\documentclass{aa}  

\usepackage{natbib}
\usepackage{graphicx}
\usepackage{caption}
\usepackage{subcaption}
\usepackage{float}
\usepackage{blindtext}

\usepackage{xspace}
\renewcommand{\hi}{H$\,$\textsc{i}\xspace} 
\usepackage{txfonts}
\RequirePackage[normalem]{ulem} 
\RequirePackage{color}\definecolor{RED}{rgb}{1,0,0}\definecolor{BLUE}{rgb}{0,0,1} 
\usepackage[pdftex]{hyperref}
\begin{document}

   \title{A comparative study of source-finding techniques in \hi emission line cubes using SoFiA, MTObjects, and supervised deep learning}


   \author{J.A. Barkai 
          \inst{1}
          \and
          M.A.W. Verheijen\inst{1}
          \and
          E. Talavera \inst{2}
          \and
          M.H.F. Wilkinson\inst{3}\fnmsep
          }

   \institute{Kapteyn Astronomical Institute, University of Groningen,
              Landleven 12, 9747 AD Groningen\\
              \email{secr@astro.rug.nl}
         \and
             Data Management \& Bionetrics (DMB), University of Twente,
             Drienerlolaan 5, 7522 NB Enschede\\
             \email{info@utwente.nl}
         \and
             Bernoulli Institute for Mathematics, Computer Science and Artificial Intelligence, University of Groningen,
             Nijenborgh 9, 9747 AG Groningen\\
             \email{bernoulli.office@rug.nl}
             }

   \date{Received September 15, 2022; accepted September 16, 2022}

 
  \abstract
   {The 21$\,$cm spectral line emission of atomic neutral hydrogen (\hi) is one of the primary wavelengths observed in radio astronomy. However, the signal is intrinsically faint and the \hi content of galaxies depends on the cosmic environment, requiring large survey volumes and survey depth to investigate the \hi Universe. As the amount of data coming from these surveys continues to increase with technological improvements, so does the need for automatic techniques for identifying and characterising \hi sources while considering the tradeoff between completeness and purity.}
   {This study aimed to find the optimal pipeline for finding and masking the most sources with the best mask quality and the fewest artefacts in 3D neutral hydrogen cubes. Various existing methods were explored, including the traditional statistical approaches and machine learning techniques, in an attempt to create a pipeline to optimally identify and mask the sources in 3D neutral hydrogen (\hi) 21 cm spectral line data cubes.}
   {Two traditional source-finding methods were tested first: the well-established \hi source-finding software SoFiA and one of the most recent, best performing optical source-finding pieces of software, MTObjects. A new supervised deep learning approach was also tested, in which a 3D convolutional neural network architecture, known as V-Net, which was originally designed for medical imaging, was used. These three source-finding methods were further improved by adding a classical machine learning classifier as a post-processing step to remove false positive detections. The pipelines were tested on \hi data cubes from the Westerbork Synthesis Radio Telescope with additional inserted mock galaxies.}
   {Following what has been learned from work in other fields, such as medical imaging, it was expected that the best pipeline would involve the V-Net network combined with a random forest classifier. This, however, was not the case: SoFiA combined with a random forest classifier provided the best results, with the V-Net--random forest combination a close second. We suspect this is due to the fact that there are many more mock sources in the training set than real sources. There is, therefore, room to improve the quality of the V-Net network with better-labelled data such that it can potentially outperform SoFiA.}
   {}


   \maketitle
%

\section{Introduction}
Large-area astronomical imaging surveys map the skies without a specific target, resulting in images that contain many astronomical objects. As the technology used to create these surveys improves with projects such as the Square Kilometre Array \citep[SKA;][]{Weltman_2020}, an unprecedented amount of data will become available. In addition to the difficulty of identifying low-intensity sources with close proximity to the level of noise, the time that it would require an experienced astronomer to manually identify sources in data sets of such volumes would be unfeasible \citep{Haigh_2021}, hence the need for fast and accurate techniques for identifying and masking sources in astronomical imaging survey data.

The 21$\,$cm spectral line emission of atomic neutral hydrogen (\hi) is one of the primary wavelengths observed in radio astronomy. It is detected via the emission of a photon due to the energy level transition of the hydrogen atom at 21cm. \hi gas is observed as clouds in galaxies, which can be used to determine their structure, as well as free-floating gas outside of galaxies (see for example \citealt{rogues} and the references therein). However, it is intrinsically faint and sensitive to its environment, requiring large volume coverage and depth to map the \hi Universe, hence the need for large redshift range surveys that use radio telescopes such as the precursors to the SKA \citep{Weltman_2020}. This includes the Meer Karoo Array Telescope \citep[MeerKAT;][]{booth2009meerkat}, a 64-dish array in South Africa, the Australian Square Kilometre Array Pathfinder \citep[ASKAP;][]{askap}, which consists of 36 dishes in Western Australia, and the APERture Tile in Focus \citep[Apertif;][]{apertif_data}, a 12-dish array upgrade of the Westerbork Synthesis Radio Telescope \citep[WSRT;][]{WSRT} in the Netherlands.

\begin{figure}
    \centering
    \includegraphics[width=0.5\hsize]{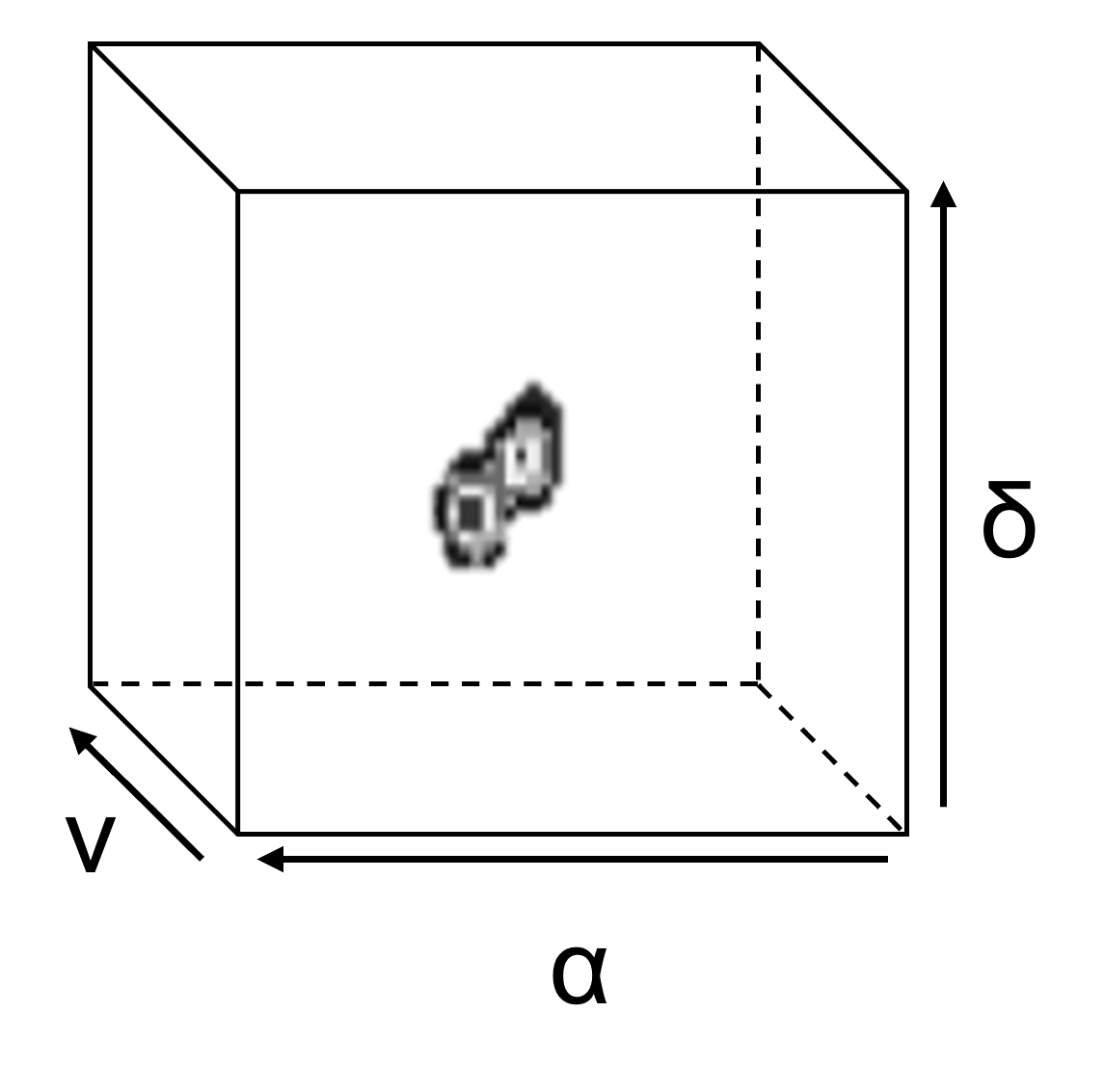}
    \caption{Schematic of a noiseless 3D \hi emission cube containing a galaxy. The cube consists of two positional dimensions (right ascension, $\alpha$, and declination, $\delta$) and one spectral dimension (frequency, $v$).}
    \label{fig:eg}
\end{figure}

The 21 cm spectral line emission of galaxies is captured and processed into floating point values whose intensities represent flux in a 3D (mosaicked) \hi data cube. The cube consists of two positional dimensions (right ascension and declination) and one spectral dimension (frequency or velocity), with a high data volume of the order of hundreds of gigavoxels \citep{Moschini2016}. A schematic of such a data cube containing an \hi emitting galaxy, without noise,  can be seen in Fig. \ref{fig:eg}. Despite the growing knowledge and tools available for source-finding in 2D images, it is still considered a challenge in astronomy, particularly when considering the same task in 3D data cubes.

Astronomical source-finding is limited by the necessary tradeoff between accepting false positives and excluding true sources. This is only heightened by the lack of well-defined boundaries and low signal-to-noise ratios of \hi sources \citep{Aniyan_2017}. In addition, most \hi sources are faint and extended \citep{punzo2015role}, and sometimes the surrounding noise that needs to be differentiated from the actual sources is very similar to the signal of the sources \citep{Gheller_2018}.

The problem at hand can be seen as an overlap between astronomy and computer vision, where existing advances in computer vision could be used to solve the struggle of finding and masking sources in astronomy. While computer vision makes use of different terminologies, such as object segmentation, the task of finding and masking astronomical sources can be defined as locating and highlighting pixels in an image that belong to different objects.

Many astronomers explore source-finding solutions using simple statistical techniques \citep{focas,sextractor,next}. However, these approaches are very sensitive to the input parameters and struggle to differentiate between faint sources and noise. The challenge lies in the absence of the boundaries of sources, with many having intensities very close to the noise, especially in the case of 21 cm data \citep{punzo2015role}. As a result, these approaches require a tradeoff between completeness and purity.

This study, therefore, aimed to address the dilemma of this tradeoff by determining the optimal source-finding pipeline, with better completeness, purity, and mask quality than previous methods. To this aim, the contributions of this work are fourfold.

First, this is the first published work, to our knowledge, that provides the training and testing of a deep learning source-finding and masking architecture on 3D \hi emission data. This model could be adapted for data from different telescopes and could be used as a source-finding method in future \hi surveys.

Second, a classical machine learning solution has been proposed for improving the purity of the catalogues created by source finders. It was trained on various methods to ensure its ability to be used on any \hi source catalogue.

Third, a new pipeline for finding and masking \hi sources in data cubes has been introduced. This was achieved by comparing statistical and deep learning methods using the well-known tool Source Finding Application \citep[SoFiA;][]{Serra_2015, sofia_2} as the baseline for the experiments. The final pipeline consists of a source-finding method followed by a machine learning model for post-processing.

Finally, a catalogue of detections by various source finders of mock \hi galaxies and their associated properties has been made available for future use\footnote{\url{https://github.com/Jbarkai/HISourceFinder.git}}. These detections were made via the various experiments and can be used for training classification models in the future.

Section \ref{chap:previous_work} begins by outlining the existing source-finding methods and an explanation of the algorithms that were tested in this paper. This is followed by an explanation of the data used and the experimental setup in Sect. \ref{chap:method}. The results of each source-finding pipeline can be found in Sect. \ref{chap:results} and are discussed and compared in Sect. \ref{chap:discussion}. Finally, some concluding statements are made and the potential for future work is explored in Sect. \ref{chap:conclusion}.
\section{Source-finding methods}
\label{chap:previous_work}
In this section, several existing astronomical source-finding methods are discussed, including their strengths and weaknesses. This is followed by the theoretical backgrounds of the source-finding software that were compared in this paper.
\subsection{Existing methods}
Existing source-finding techniques can be categorised into three types of approaches. The first is traditional methods, which use statistical techniques to identify sources, followed by classical machine learning techniques and deep learning approaches. Since most source-finding techniques are originally designed for 2D images, pixels will often be mentioned, but it is important to note that 3D data cubes are made up of voxels.
\subsubsection{Traditional methods}
The two most common traditional source-finding techniques in astronomy are thresholding and local peak search \citep{10.1111/j.1365-2966.2012.20742.x}. The simplest of the two is thresholding, where a collection of connected pixels are masked as a source if they are above a specified threshold value. The threshold can be global to the image but due to background variations it is often chosen locally, depending on the neighbourhood, and in some cases, an adaptive threshold is used. However, despite their high speeds and simplicity, thresholding methods have difficulty finding faint sources. SExtractor \citep{sextractor} is one of the most well-known source-finding programmes in astronomy that uses a thresholding method.

The other most widely used traditional method is local peak search, which finds sources by creating a list of candidate pixels for the central points of sources, chosen by the local maxima. The sources are then masked by taking the pixels around each central point candidate that decrease in intensity. While local peak search has already been in use since the late 1970s \citep{Argueso_2006}, it is more suited for detecting point sources and struggles to find extended sources. A common weakness of both thresholding and local peak search methods is their difficulty in de-blending or differentiating between multiple sources \citep{Haigh_2021}.

While thresholding and local peak search are the most common classical source-finding techniques developed, in recent years a variety of methods have been applied, inspired by other computer vision applications. A popular classical source-finding technique in computer vision is edge-based source-finding and masking. This algorithm detects the edges of a source by finding discontinuities in the characteristics of the pixels, such as the brightness. These edges are then connected to create a border around the mask of the located source \citep{edgeseg}. However, this approach is more appropriate for objects with well-defined edges and therefore is not suitable for the detection of diffuse \hi emission sources.

Another popular classical source-finding method is the watershed transform \citep{2018arXiv181003908B}. This method treats the image as a topographic surface and interprets the gradient of the pixel intensity as the elevation. It then locates and masks sources in the image by acting as if water is poured into the minima and treats the boundaries as where the puddles would merge. This is used for source-finding in astronomy by \citet{7071090}, resulting in masks with continuous boundaries. Although the watershed approach is more stable than edge-based methods, it suffers from its computational complexity \citep{Kaur2014VariousIS}.

Region-based detection locates sources and creates masks by grouping pixels with similar characteristics. While these region-based methods are popular among computer vision tasks, they have only recently been introduced in the context of astronomical source-finding. These methods have the advantage of being less sensitive to noise, particularly when prior knowledge can be used to define which characteristics to compare the pixels. However, region-based methods are also known to be memory and time-intensive \citep{survey_segmentation}. A recent example of region-based source-finding software used in astronomy is Max-Tree Object (MTObjects or MTO) \citep{Teeninga,Teeninga2016}, which makes use of max-trees (see Sect. \ref{sec:MTO}). This software was originally designed for 2D optical data. However, since diffuse optical sources similarly suffer from the tradeoff between completeness and purity, MTO has been extended further for \hi emission cubes by \citet{Arnoldus2015}.

Weighing up the pros and cons of each of these methods, the best source-finding solution, therefore, depends on the astronomical data itself. Hence the creation of general source-finding software, the flexible SoFiA \citep{Serra_2015, sofia_2}. SoFiA is designed to be independent of the source of \hi emission line data used and is currently the most used pipeline for source-finding in \hi emission cubes. It was therefore used as the benchmark for our experiments.
\subsubsection{Classical machine learning methods}
\label{sec:ml}
In addition to the traditional statistical methods used for source-finding, there are many machine learning approaches. Classical machine learning methods learn from the characteristics of the observational data, such as the brightness of a pixel, also known as features. On the other hand, deep learning takes the raw data itself as input and then determines the features. These features are chosen to represent the physical properties of the system and therefore their quality determines the efficiency of the learning algorithm. It is therefore important to have a good understanding of the data being used, and in this case radio astronomy, to implement these techniques.

Both classical and deep machine learning methods can be either supervised or unsupervised. Unsupervised methods find the boundaries of objects using their intensity or a gradient analysis, which makes them better suited for identifying sources with well-defined boundaries. Supervised methods instead use prior knowledge through training samples. Since the \hi sources in the data cubes are known to have ill-defined boundaries \citep{punzo2015role}, supervised machine learning techniques are better suited.

The most well-researched techniques use classification as a way to find sources, including support vector machines \citep{SupportV79}, K-nearest neighbours \citep{nearestneighbour}, decision trees \citep{decisiontree}, and random forests \citep{598994}. While these techniques were not used directly as source finders, the random forest \citep{598994} classifier was used as a post-processor to improve the purity of the catalogues created by the source-finding methods, since it is one of the best-performing algorithms for classification \citep{RF_best}.

The random forest \citep{598994} algorithm, as its name suggests, creates a random forest of uncorrelated decision trees \citep{decisiontree}, which essentially build trees based on the features of a data set. To reduce variance, the decision trees are each trained on a subset of the training data and input features are selected randomly with replacement, also known as bootstrapping. The bias in the learning error is reduced by merging the decision trees and taking the average result or the result with the most votes. Random forests are therefore advantageous in their accuracy, stability, and few tuning parameters, but they can be very slow compared to other classical machine learning classifiers if a large number of trees are used \citep{RF_best}.
\subsubsection{Deep learning methods}
Deep learning is a type of machine learning that learns directly from observational data rather than its features. In recent years it has been the primary method used for finding and masking sources in computer vision due to its high accuracy above traditional methods \citep{RF_best}. Since deep learning takes the raw data as its input, it requires less prior knowledge of the data for training and therefore makes it most advantageous when the features do not provide enough information to create an accurate model. The main drawbacks of deep learning methods are their lack of transparency, their computational expense, and the need for large training data sets with already located and masked sources \citep{DBLP}. However, there is already a lot of work done to overcome the latter two challenges \citep{samudre2020dataefficient, ZhiLLG18, bochkovskiy2020yolov4}.

The most frequently used deep learning method for image processing is the convolutional neural network \citep[CNN;][]{CNNs}. A CNN is a type of artificial neural network that contains one or more convolutional layers. An advantage of CNNs is their ability to be re-trained on custom data sets, which allows for more flexibility while classical machine learning methods tend to be more domain-specific \citep{DBLP}. It is not only widely used in the field of computer vision but has recently become a popular method for source-finding in 21 cm astronomy images \citep{Gheller_2018, lukic2019convosource, Aniyan_2017}, and is found to out-perform other machine learning methods when used for optical 2D galaxy classification \citep{Cheng_2020, Alhassan_2018}.

While the use of CNNs for 2D images is well explored, the application of CNNs on 3D data cubes is still very new. The main challenge is that CNNs require a fixed-dimension grid input. Many of the existing methods use 2D data to train the network due to the lack of 3D data  containing masks available for training. Many different techniques exist that essentially slice the cube into multiple 2D slices and then fuse them back together to create a masked cube, for example, \citet{3d_unet} and \citet{Yang2021}. The identification of sources in each slice is done based on the top-ranking medical imaging CNN architecture called U-Net \citep{ronneberger2015unet}.

Applying 2D CNNs to slices of the data has the advantage of being able to make use of previously trained weights from studies such as \citet{Gheller_2018} and \citet{lukic2019convosource}, which find sources in 2D radio data. However, while this saves computational costs, it is very important that the data are sliced in a way that preserves the shape of the sources, which is not so simple with asymmetric sources. An alternative approach is using an architecture that takes a 3D volume as its input, as opposed to 2D slices, and uses volumetric convolutions. V-Net \citep{Milletari2016} is an architecture that does just that, following the architecture of U-Net \citep{ronneberger2015unet}.

\subsection{Comparing source-finding software}
While there are a variety of source-finding techniques that exist, the traditional methods require extensive prior knowledge and still have been found to fail at differentiating many faint \hi sources from surrounding noise. Although both classical machine learning methods and even deep learning have begun to be explored in astronomical source-finding, there is very little work done that uses these techniques on 3D \hi emission data. In addition, the 3D CNN architectures that do exist have not yet been tested for finding astronomical sources. We, therefore, investigated the use of a 3D CNN architecture, namely V-Net \citep{Milletari2016}, to locate and mask \hi sources in 3D data cubes. This was then evaluated and compared to software that makes use of traditional methods, SoFiA \citep{Serra_2015, sofia_2} and MTO \citep{Teeninga2016}, which will be discussed first.

\subsubsection{SoFiA}
Comparing the strengths of the best traditional algorithms used for source-finding, it is clear that the optimal source finder depends on the properties of the astronomical data itself. This has led to the creation of general source-finding software, the flexible SoFiA \citep{Serra_2015, sofia_2}. SoFiA enables the choice of a variety of source-finding techniques and has been designed to be independent of the type of \hi emission line data used. Currently, this is the most used pipeline for source-finding in \hi emission data and is, therefore, the benchmark for the experiments for this study. The pipeline begins by removing any variation in the noise to ensure the assumption of uniform noise. This is done by normalising the cube with a weights cube made up of the local noise.

Version 2.3.1 of SoFiA \citep{Serra_2015, sofia_2} makes use of the smooth and clip method since this technique has been found to be the best source-finding method for extended sources \citep{Popping_2012} in \hi data cubes. This algorithm uses 3D kernels to iteratively smooth the cube and find sources on multiple scales, both spatially and spectrally. For each resolution, voxels are added to the mask on the condition that they have a flux relative to the global root mean square (RMS) noise level above the chosen signal-to-noise threshold. Since \hi sources are known to have an exponential radial surface brightness profile \citep{hi_profile} and a double-horned spectral profile, a Gaussian filter is used in the spatial domain and a boxcar filter in the spectral domain. 

Once the voxels associated with sources are identified, an algorithm equivalent to the friends-of-friends algorithm \citep{CHNI} is used to merge them into sources. This is done by looping over the segmented binary mask and assigning a label to unlabelled source-containing voxels as well as the neighbouring voxels within a chosen merging or linking length. In this process, sources that do not fit the given size criteria are rejected.

The reliability of the sources is then determined following \citet{serra_jurek_2012a}. This works by assuming that a true source is positive and that the noise in the cube is symmetric. With this assumption, it is implied that there is an equal number of noise peaks with a negative intensity as there are with a positive intensity. The reliability of a source can therefore be evaluated in a 3D parameter space by measuring the density of positive and negative sources and calculating the probability that a positive source is not just a noise peak. Detections that are found to have an integrated signal-to-noise (S/N) ratio below an assigned S/N limit are designated a reliability score of zero. The resulting reliability score can be thresholded to remove unreliable sources.
\subsubsection{MTObjects}
\label{sec:MTO}
MTObjects \citep{Teeninga,Teeninga2016} is region-based source-finding software that makes use of max-trees. Max-trees are a special case of a component tree, a tree built from connected components of threshold sets. Connected components are classically defined as path-connected groups of voxels with the same intensity, foreground or background in the thresholded case, of maximal extent \citep{nagel}. The definition of a max-tree, according to \citet{663500}, is a representation of an image as a tree structure with the maxima as its leaves. The root of the max-tree consists of the entire image domain, while the leaves represent the voxels with the local maxima values. The nodes of the max-tree represent connected components of threshold sets of the image, where it is the grey scale of the image that is thresholded. Therefore, there can never be more nodes than there are voxels in the image. Each connected component of the max-tree can be assigned attributes such as the volume and flux density, which allows for filtering based on a threshold for these attributes.

\begin{figure}
    \centering
    \includegraphics[width=0.7\hsize]{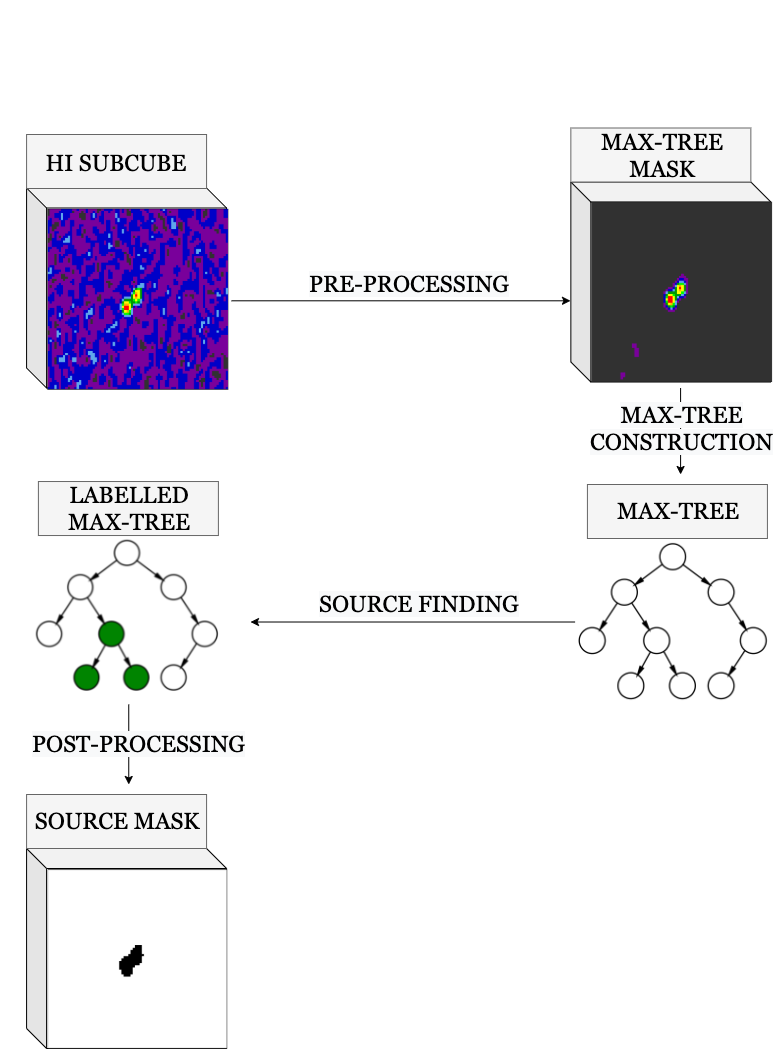}
    \caption{Schematic diagram of the source-finding pipeline of MTObjects \citep{Arnoldus2015}. An \hi emission sub-cube, which contains a source, is transformed into a max-tree mask with the use of both Gaussian and adaptive smoothing. A max-tree is then built from the max-tree mask, and the source nodes are labelled using the attributes of the smoothed cube. Finally, post-processing is used to remove false detections.}
    \label{fig:mtoflow}
\end{figure}

A max-tree is much more compact than a classical component tree, saving memory by not storing the elements of their children and consisting of only those component tree nodes that have at least one voxel at the appropriate threshold level. The advantage of max-trees is that they can filter the cubes without distorting any edge information because it operates on the connected components of an image rather than individual voxels \citep{663500}. The version of MTObjects implemented by \citet{Arnoldus2015} was used, as it was designed for 3D radio cubes. The pipeline consists of four steps that can be seen in the flowchart in Fig. \ref{fig:mtoflow}.

The first step of the MTO pipeline is the pre-processing of the cube to create a max-tree mask. First, the cube is smoothed with a Gaussian filter to stabilise the gradient approximation. This is followed by the application of spatially adaptive smoothing to ensure the preservation of the edges of sources. The adaptive smoothing technique used is known as the Perona–Malik diffusion filter \citep{56205}. The filter models the smoothing as an anisotropic diffusion process.

Both smoothing steps are used to increase the signal-to-noise ratio and are followed by a background estimation to create a max-tree mask. The background is calculated as the mean of the regions that contain no sources. The background is then subtracted from the smoothed image and the resulting negative values are set to zero to ensure the lack of negative-total-flux nodes in the max-tree.

Once the cube has been smoothed and the background subtracted, the next step of the pipeline is the construction of the max-tree. During the construction of the max-tree, the node attributes are also calculated to inform the removal of false detections at a later stage.

Due to the limited classical definition of connectivity, \cite{db03eb7e637742658eb53afef596821d} presented a new approach known as mask-based connectivity that instead relies on the pre-processed connectivity mask. However, \citet{Arnoldus2015} adapts this definition of connectivity further still by storing the full original volume to calculate component attributes or properties at a later stage in the max-tree data structure. This allows for the enhancement of its source-finding by using the pre-processed image while maintaining the original image for the statistical model.

Once the max-tree is constructed, MTO finds and labels the nodes that belong to a source, this is the object detection step in Fig. \ref{fig:mtoflow}, which filters the max-tree and chooses the nodes based on their attributes using statistical tests. Since the simple $\chi^2$ method from \citet{Teeninga2016} would not suffice in modelling the correlated noise in the \hi cubes, the statistical tests have been modified for radio data \citep{Teeninga}. The attribute used to detect sources was chosen to be the flux density, as it is less sensitive to outliers \citep{Arnoldus2015}. The flux density is defined as follows:
\begin{align}
    F_{\mathrm{density}}(P) &= \frac{F_{\mathrm{total}}(P)}{V(P)}, \\
    &= \frac{1}{V(P)}\sum_{x\epsilon P}f(x),
\end{align}
where $P$ is a peak component of the max-tree mask, $V(P)$ is its volume, and $f$ is the original volume. A node or part of the volume is included in a source mask if it has a greater value of this attribute than other children of its parent node or if it has no significant parent.

\subsubsection{V-Net}
\label{sec:method_vnet}
As mentioned, the most used deep learning method for image processing is the CNN. While CNNs can be designed from scratch, there are many already existing architectures that have been optimised for source-finding tasks \citep{FCN, Christ2016, Milletari2016}. These architectures can be adapted to a variety of source-finding tasks and data sets, and can either be re-trained from scratch or the weights from previous tasks can be selectively updated.

The CNN architecture chosen for this project was the novel V-Net \citep{Milletari2016} network. V-Net is originally designed for locating and masking objects in medical images \citep{Milletari2016} and was chosen due to its ability to take full data volumes as its input as opposed to slices of 3D images. V-Net is a fully volumetric CNN built following the well-known architecture of U-Net \citep{Christ2016}. 

A schematic diagram of the V-Net architecture is shown in Fig. \ref{fig:vnet}. Following Fig. \ref{fig:vnet} left to right, the model is composed of a compression and decompression path, much like U-Net. Each path is composed of multiple stages containing one to three convolutional layers each, which enable the learning of a residual function that ensures convergence, unlike U-Net. The convolutional layers in the stages of the compression path have a volumetric kernel of $5\times5\times5$ voxels. In between each stage, a $2\times2\times2$ voxels filter is applied with a stride of two voxels to halve the spatial sizes of the output feature maps and double the number of feature channels. This step is followed by the use of the Parametric Rectified Linear Unit \citep[PReLU;][]{2015arXiv150201852H} as an activation function. This is a modification of the well-known Rectified Linear Unit \citep[ReLU;][]{agarap2018deep}, which directly outputs the input if it is positive and otherwise outputs zero. PReLU simply generalises this by using a slope for negative values.

The decompression path increases the spatial size of the input of each layer using de-convolutions, which are followed by one to three convolutional layers with filters the same size as those used in the compression path. With each convolutional layer, the number of filters is doubled and again a residual function is learnt. As a result, two output feature maps are produced with the same dimensions as the volume inputted into the network. These two probabilistic maps of the foreground and background regions are produced using voxel-wise softmax regression \citep{2017arXiv170400805G}, which normalises the map so that the voxel values follow a probability distribution. As in U-Net, horizontal links are used at each stage to propagate the extracted features to the decompression paths in order to recover the location information lost in the compression path.

\begin{figure}
    \centering
    \includegraphics[width=0.7\hsize]{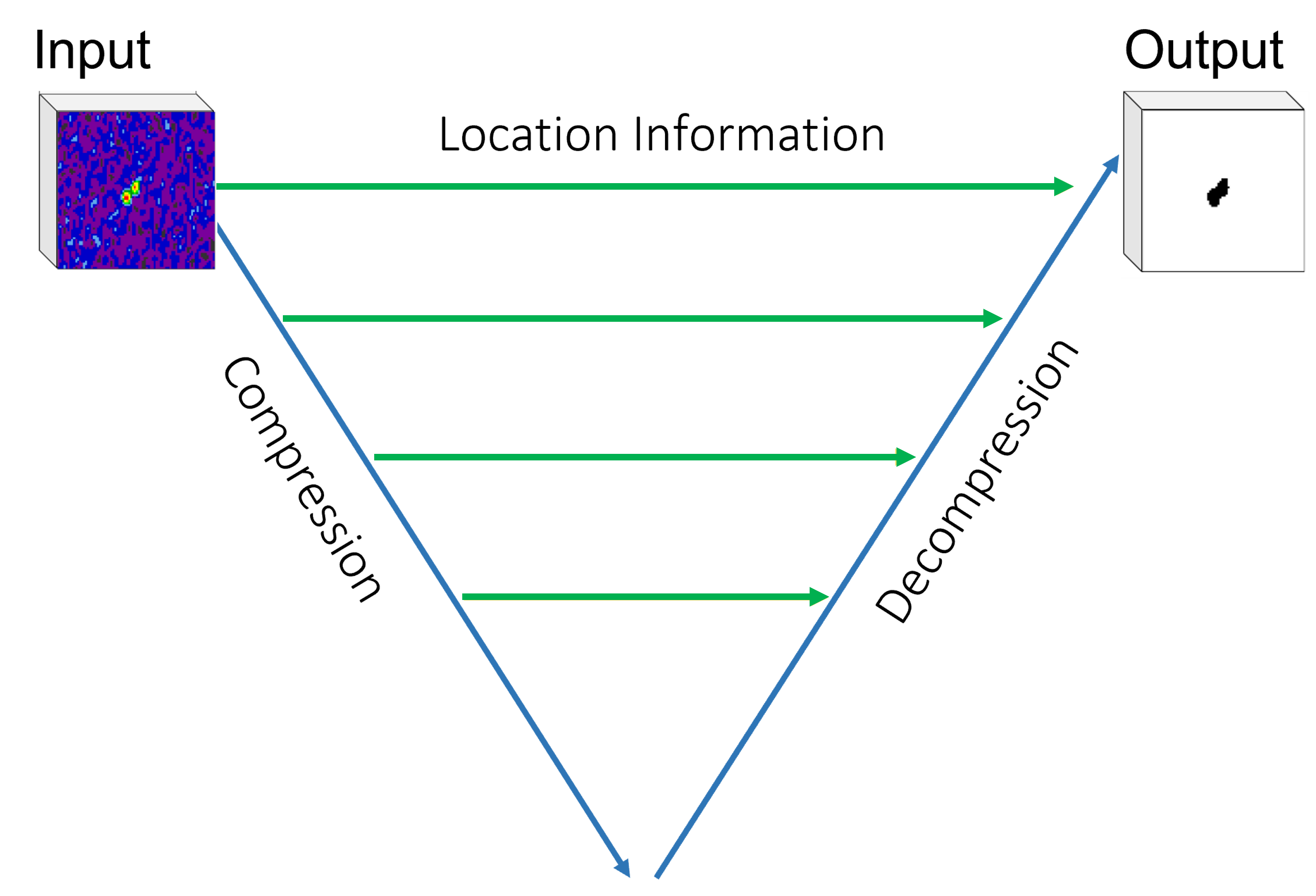}
    \caption{Illustration of the V-Net architecture, based on the architecture of \citet{Milletari2016}. The compression path is shown on the left and the decompression path on the right, and the two are horizontally linked by the location information.}
    \label{fig:vnet}
\end{figure}

The sources aimed to be found occupy only a very small region of the volume, which only increases the risk of the training process getting stuck in a local minimum of the loss function. This could result in predictions that are biased to the noise and the model may miss or partially detect sources. Since this is an issue for medical imaging \citep{reweight}, sample re-weighting-based loss functions can be employed \citep{FCN, Christ2016}. These types of functions favour the sources during the learning phase. However, another approach to counter-act this imbalance is to use a loss function and evaluation metric that is inherently balanced. The most common of these is the Dice coefficient \citep{dice_coefficient2, dice_coefficient}, also known as the F-score \citep{fscore}. The F-score maximises the measure of the overlap of two masks, ranging from 0 to 1. With $A$ and $B$ being two overlapping masks, the F-score is defined as twice the intersection divided by the union of each mask as follows:
\begin{align}
    \label{eq:fscore}
    D &= F_{score} = \frac{2 (A\cap B)}{A + B}, \\
    &= 2\times \frac{\mathrm{precision} \times \mathrm{recall}}{\mathrm{precision} + \mathrm{recall}},
\end{align}
where the recall or the completeness measures the proportion of detected sources and the precision or purity measures the proportion of segments that match the ground truth or target masks as follows:
\begin{align}
\mathrm{precision} &={\frac{tp}{tp+fp}},\\
\mathrm{recall} &={\frac {tp}{tp+fn}}.
\end{align}
Here $tp$, $tn$, $fp$, and $fn$ are the true positives, true negatives, false positives, and false negatives, respectively. An adaption of the Dice coefficient is used for the loss function of the V-Net network and is defined as follows \citep{Milletari2016}:
\begin{align}
    D = \frac{2 \sum^{N}_{i} p_ig_i}{\sum^{N}_{i} p_i^2 + \sum^{N}_{i} g_i^2},
\end{align}
where $N$ is the number of voxels in the detection mask, $p_i$, with the ground truth mask $g_i$. The gradient with respect to the $j^{th}$ voxel is therefore defined as\begin{align}
    \frac{\partial D}{\partial p_j} = 2\Bigg{[}\frac{g_j(\sum^{N}_{i} p_i^2+\sum^{N}_{i} g_i^2)-2p_j(\sum^{N}_{i} p_ig_i)}{(\sum^{N}_{i} p_i^2+\sum^{N}_{i} g_i^2)^2}\Bigg{]}.
\end{align}
This allows a balance between the foreground and background voxels without the need to assign them weights. Moreover, this loss function is shown to outperform those with sample re-weighting methods \citep{Milletari2016}.
\section{Methodology}
\label{chap:method}
This section describes both the observed and synthetic data used to compare the source-finding methods described in the previous section. This is followed by an explanation of the experimental setup, including the machines used for the experiments and the parameters used for each source-finding software.

\subsection{Data}
\label{chap:materials}
Supervised machine learning methods, especially CNNs, require a large sample of labelled training data, which is not easily available for radio images. Previous approaches, such as Galaxy Zoo \citep{Lintott_2008}, make use of crowdsourcing to create manually labelled data via visual inspection. However, even if manually labelled data were available, the ground truth or target mask is not known and some labels could be missed or incorrect, due to the difficulty of interpreting the 3D data cubes. Additionally, many of the sources of interest are rare and are therefore very difficult to train to detect \citep{Haigh_2021}. To tackle this problem simulated cubes of \hi emission containing galaxies with labelled masks were used for this project.

\subsubsection{Observed data}
To prevent the need to simulate noise, existing observed data cubes were taken from the \hi medium-deep imaging survey from Apertif \citep{ApertifS94} receivers on the WSRT \citep{WSRT}. Apertif collects data from 40 beams on the sky simultaneously, which are then combined into a single mosaic data cube. Each of the individual, nearly circular, beams covers 36$'$ of the sky, resulting in the final mosaic covering about 6.2 square degrees. The observations cover 1279.9$\,$MHz to 1425.1$\,$MHz, which are then split into seven spectral windows of 23.84$\,$Hz, for parallel computation, each with 652 channels. These windows were split in such a way that they overlap in frequency to account for the frequency width of galaxies.

\begin{figure}
     \centering
         \includegraphics[width=\hsize]{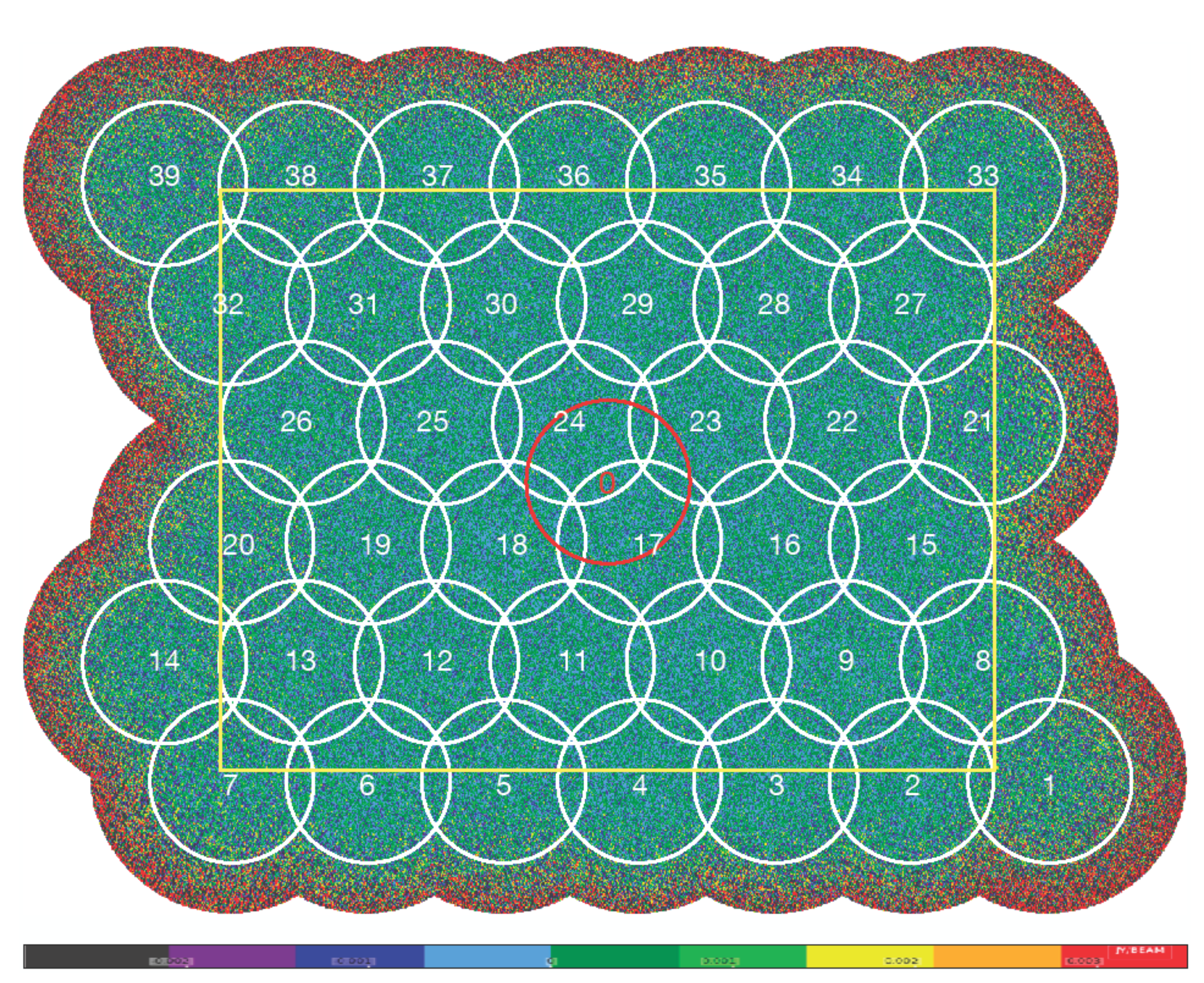}
        \caption{Example of a single channel of an Apertif \hi emission cube, prior to noise normalisation. The colour scale represents the noise balanced on 0, which is non-uniform and increases towards the edges of the cube. The layout of the Apertif compound beams mosaicked into a hexagonal shape is overlaid in white, following \citet{apertif_data}. Note that the cubes were cropped spatially, here indicated by the yellow rectangle, resulting in $1000\times1600$ pixels in dimension, representing the declination and right ascension, respectively.}
        \label{fig:apertif_eg}
\end{figure}

Not every beam is equally sensitive and therefore the observations contain significant variations in the noise. In order to correct for these noise variations and allow for better source-finding, the noise was normalised by dividing the \hi emission cube by the spectral RMS noise at each pixel. Figure \ref{fig:apertif_eg} shows both the layout of the beams that form the mosaic and an example of a single channel taken from one of the data cubes prior to its noise normalisation. Due to the high intensity of noise at the edges and the blank voxels in the corners of the \hi emission cubes, the corners and edges were cropped to prevent any errors from the source finders in case they could not take blank voxels in their input.

The observed areas of the sky in the imaging surveys performed with Apertif were chosen to include regions with publicly available spectroscopic and optical observations. Of the two major imaging surveys, the data was taken from the medium-deep survey \citep{ApertifS94}, which covers an area including the Perseus–Pisces Supercluster \citep{B_hringer_2021}. This area consists of nine pointings, each observed 10 times for 11.5 hours to reduce the noise by a factor of $\sqrt{10}$ and increase the sensitivity when combining them. Due to the partial completion of the observations at the start of this project, two out of the nine pointings were used, one for training purposes and one for testing, each containing six spectral windows.  The pointing used for training purposes ranges from $1^{\mathrm{h}}45^{\mathrm{m}}10.92^{\mathrm{s}}$ - $2^{\mathrm{h}}5^{\mathrm{m}}26.57^{\mathrm{s}}$ in right ascension and $34^{\circ}51'04.49''$ - $37^{\circ}50'56.31''$ in declination with a resolution of 25$''$ in the north-south direction. The pointing used for testing purposes covers $1^{\mathrm{h}}45^{\mathrm{m}}29.82^{\mathrm{s}}$ - $2^{\mathrm{h}}5^{\mathrm{m}}7.69^{\mathrm{s}}$ in right ascension and $32^{\circ}25'16.19''$ - $35^{\circ}25'08.58''$ in declination with a resolution of 27$''$ in the north-south direction. At the rest frequency of 1420$\,$MHz, applying a robust weighting of zero, the angular resolution in the east-west direction for both pointings is 15$''$.

Once cropped, each of the spectral windows from each pointing makes up a data cube with $652\times1000\times1600$ voxels in dimension, representing the frequency, declination and right ascension, respectively. The angular size of the pixels is 6$''$ and the spectral width of the channels is $36.621\,\mathrm{kHz}$. The properties of the six spectral windows for the two pointings can be found in Table \ref{tab:apertif_properties}. The RMS of the noise prior to normalisation shows how the noise varies with the frequency range, increasing for more distant spectral windows.

\begin{table*}[ht]
    \centering
    \caption{Properties of the seven spectral windows of the two pointings chosen from the Apertif medium-deep survey.}
    \begin{tabular}{cccccc} \\ \hline 
         & Frequency Range & Redshift & Distance & Recession Velocity & RMS noise \\
         & (MHz) & Range & Range (Mpc) & Range (km/s) & (mJy/Beam) \\ \hline \hline 
        Training Pointing & 1380.96 - 1404.80 & 0.028561 - 0.011106 & 122 - 48  & 8562 - 3329 & 1.69 \\
        & 1360.75 - 1384.59 & 0.043841 - 0.025868  & 188 - 111  & 13143 - 7755 & 1.71 \\
        & 1340.53 - 1364.38 & 0.059582 - 0.041067 & 255 - 176  & 17862 - 12312 & 1.72 \\
        & 1320.32 - 1344.16 & 0.075804 - 0.056724 & 325 - 243  & 22726 - 17005 & 1.73 \\
        & 1300.10 - 1323.95 & 0.092532 - 0.072858 & 396 - 312  & 27740 - 21842 & 1.77 \\
        & 1279.89 - 1303.73 & 0.109787 - 0.089493 & 470 - 383  & 32913 - 26829 & 2.17        \\
        Test Pointing & 1381.04 - 1404.88 & 0.028506 - 0.011053 & 122 - 47  & 8546 - 3314 & 1.46 \\
        & 1360.82 - 1384.66 & 0.043785 - 0.025813 & 188 - 111  & 13126 - 7739 & 1.49 \\
        & 1340.61 - 1364.45 & 0.059524 - 0.041011 & 255 - 176  & 17845 - 12295 & 1.51 \\
        & 1320.39 - 1344.23 & 0.075745 - 0.056666 & 324 - 243  & 22708 - 16988 & 1.53 \\
        & 1300.18 - 1324.02 & 0.09247 - 0.072799 & 396 - 312  & 27722 - 21825 & 1.91 \\
        & 1279.96 - 1303.80 & 0.109724 - 0.089432 & 470 - 383  &  32894 -26811 & 2.43 \\
        & &  &  &  &  \\
        \hline
    \end{tabular}
    \label{tab:apertif_properties}
\end{table*}
\subsubsection{Creating and inserting the mock galaxies}
\label{sec:insert}
The 3000 mock galaxies created by \citet{Gogate2022} were used as a parent library for the insertion into the \hi emission cubes, similar to that shown in Fig. \ref{fig:apertif_eg}. These galaxies were created following empirical scaling relations and by assuming a $\Lambda$ cold dark matter cosmology with $\mathrm{H}_0=70 \,\mathrm{km}\, \mathrm{s}^{-1}\,\mathrm{Mpc}^{-1}$ and $\Omega_{\mathrm{m}}=0.3$ using Galaxy Model (GalMod), which is part of the Groningen Image Processing SYstem \citep[GIPSY;][]{VanderHulst1992}. In order to create the mock galaxies, GalMod requires a rotation curve and a radial \hi density distribution for each galaxy. The universal rotation curve prescription by \citet{1996MNRAS.281...27P} was used to calculate the rotational velocities of the mock sources. The radial \hi surface density distributions were based on the prescription by \citet{2016A&A...585A..99M} and slightly modified by \citet{Gogate2022}. We refer to \citet{Gogate2022} for further details.

From the available library of 3000 mock galaxy cubes, 300 were randomly selected and inserted into each of the twelve Apertif data cubes, resulting in a total of 1800 mock galaxies inserted in the six training cubes, and 1800 inserted in the six test cubes. The insertion of the mock galaxies into the observed Apertif cubes guarantees that the CNN was trained to find sources in the presence of noise with realistic properties.

As mentioned, the mock galaxies were scaled in size and flux as if observed at a distance of 50$\,$Mpc, corresponding to a redshift of z=0.011675, assuming a quiet Hubble flow with H$_0=70\,$km$\,$s$^{-1}\,$Mpc$^{-1}$, and an observed frequency of 1404.014 MHz. The mock galaxy cubes have pixels of 5$''$ or 1.212 kpc, and a channel width of 24.206 kHz or a rest frame velocity width of 5.169 km$\,$s$^{-1}$. Furthermore, the cubes were smoothed to mimic a Gaussian synthesised beam with an angular resolution of $\Theta=$15$''\times$25$''$ corresponding to 3.6$\times$6.1 kpc$^2$.

Inserting a mock galaxy into one of the Apertif cubes that were corrected for primary beam attenuation, consisted of multiple preparatory steps. First, a random position and channel were chosen. The frequency, f$_{obs}$, of the chosen channel corresponds to a redshift of z$_{obs}$=(f$_{rest}$/f$_{obs}$)$-$1, where f$_{rest}$$=$1420.405 MHz is the rest frequency of the \hi\ emission line, which in turn corresponds to a distance of D$_{obs}$$=$cz/H$_0$ Mpc, where c=299792.458 km$\,$s$^{-1}$ is the speed of light. It should be noted that since the angular resolution of the pre-existing mock galaxy cubes can only be reduced and not improved, the observing frequency corresponding to the distance of 50 Mpc was also the maximum frequency at which the mock galaxies could be inserted into the Apertif cubes.

Second, the angular resolution of the mock galaxy cube was reduced by smoothing to a Gaussian synthesised beam of $\Theta_{obs}=$ (1+z$_{obs}$)15$''$$\times$(1+z$_{obs}$)25$''$, after which the channels were re-gridded to pixels of size (1+z$_{obs}$)$\,$6$''$. Third, the mock galaxy cube was re-sampled in the spectral direction to a channel width of 36.621 kHz, while accounting for the corresponding rest frame channel width in km$\,$s$^{-1}$, at the observing frequency f$_{obs}$. Fourth, the flux density of the signal in the mock galaxy cube was scaled to the distance, D$_{obs}$, and beam size, $\Theta_{obs}$, such that the measured total \hi mass of the mock galaxy would be preserved.

\begin{figure}
    \centering
    \includegraphics[width=\hsize]{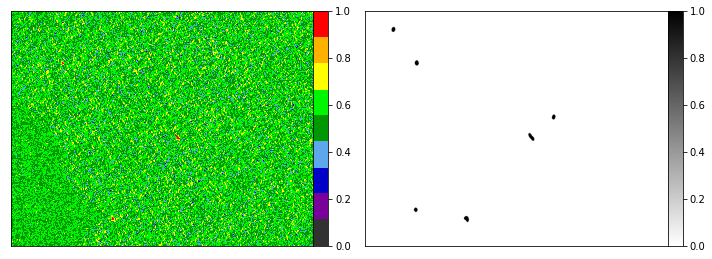}
    (a)\quad \quad \quad\quad \quad \quad \quad\quad \quad \quad\quad \quad(b)
    \caption{Zoomed-in, z-scaled \citep{Advanced27:online} single channel of an example of an Apertif cube (a) and the corresponding binary mask of the inserted mock galaxies (b). The cube has been noise-normalised, and mock galaxies have been inserted. The colour bars represent the intensity of the simulated emission.  }
    \label{fig:loud}
\end{figure}

Finally, for the purpose of labelling the ground truth in the training process for the 3D CNN and for evaluating all three source-finding methods, a binary mask cube was also made for each mock galaxy cube by converting all positive flux voxels to one and the rest of the voxels to zero. These binary mask cubes were then inserted into a duplicated Apertif cube in which all the voxels were first set to zero. Figure \ref{fig:loud} illustrates a channel from an Apertif data cube after insertion of the mock galaxies, along with the corresponding binary mask.

The smoothing of the mock galaxy cubes, the resampling of the voxels, and the scaling of the flux allowed a voxel-by-voxel insertion of the mock galaxy cube into the Apertif data cube around the randomly selected channel. It should be noted, however, that the resulting data cubes with realistic noise properties not only contain the 300 inserted mock galaxies but also dozens of unknown and unlabelled real sources. Section \ref{sec:real_sources} discusses how these real sources were identified and treated in the training process.

\subsection{Experimental setup}
\label{chap:experiment}

All three source-finding algorithms were executed on the same compute architecture, a single machine consisting of two Intel Xeon E5-2698 v3 central processing units (CPU), each with 16 cores and 32 threads, and equipped with 24$\times$32$=$768 GB DDR4 2133 MT/s of RAM. The training of the V-Net network, however, was done using the Peregrine high-performance computing (HPC) cluster of the University of Groningen \footnote{\protect{\url{https://wiki.hpc.rug.nl/peregrine/introduction/cluster_description\#dokuwiki__top}}}, consisting of 5740 CPU cores and 220000 compute unified device architecture (CUDA) cores \citep{cuda}. The training of the V-Net network used only one of these CUDA nodes with an allocated memory of 9GB.

Version 2.3.1 of the SoFiA software \citep{Serra_2015, sofia_2} was used for this experiment. This version was written in C, making it much faster than previous versions implemented in Python. For MTObjects, the code developed by \citet{Arnoldus2015} was used, since it is designed for 3D radio data cubes. The V-Net network was implemented in PyTorch \citep{NEURIPS2019_9015} and is available through MedicalZooPytorch\footnote{\url{https://github.com/black0017/MedicalZooPytorch}} \citep{adaloglou2019MRIsegmentation}, an open-source 3D medical segmentation library. PyTorch is an open-source machine learning Python library that supports CUDA tensor types. It can therefore utilise graphics processing units (GPU) for computational purposes, which is advantageous for computer vision tasks.

The details of the parameters used for the implementation of SoFiA, MTO, and the V-Net network can be found in the Appendix. The move-up factor and the connectivity of MTO were chosen following \citet{Arnoldus2015}. The smoothing kernels used for SoFiA were chosen based on astrophysical considerations, such as the expected physical size and line width of galaxies (for example \citet{sofia_2}). In addition, the minimum spatial and spectral widths for both MTO and SoFiA were chosen based on the mask sizes of the mock galaxies. The parameters of MTO and SoFiA were then manually optimised for the Apertif data cubes, by visually inspecting the resulting masks and comparing them to the ground truth masks of only the mock galaxies such that the parameters were optimised for purity without compromising the completeness.

The V-Net network, as with other neural networks, makes use of an optimiser to automatically modify its parameters to reduce its loss score. Using the adaption of the Dice coefficient as the loss function, explained in Sect. \ref{sec:method_vnet}, the loss score is calculated by comparing the predictions to the targets and minimising the loss function with each epoch. For the implementation of V-Net, the adaptive moment estimation \citep{kingma2017adam} was chosen as the optimiser because it combines the best properties of resilient backpropagation \citep{Igel00improvingthe} with those of another optimiser called an adaptive learning rate method. It has also been found to outperform the well-known stochastic gradient descent \citep{ruder2016overview} when many parameters are used and when saddle points are encountered. Through experimentation, a learning rate of 1$\times$10$^{-3}$ was chosen to minimize the loss score without drastically slowing down the learning process. 

Prior to testing the source finders, V-Net was trained using a ground truth containing the masks of only the mock galaxies, since the location of real sources in the data cubes was not yet known. All three source-finding methods were run on the six Apertif cubes of each of the two pointings, containing real sources, realistic noise and the inserted mock galaxies. The V-Net network expects an input of dimensions $128\times128\times64$ voxels, with the channels as the last dimension as opposed to the first. Therefore, a sliding window was used to create the input for the network with an overlap of size $15\times20\times20$ voxels. Due to the high memory usage of MTO, a sliding window was also used to run the source finder on smaller inputs. Since the size of the windows was only limited by the memory of the machine used, the largest windows possible were taken. The windows therefore each consisted of $652\times200\times300$ channels and pixels, respectively.

Each method yielded a catalogue of sources consisting of mock galaxies, potential real sources and false positives. Mock galaxies in the catalogue were identified by cross-matching against the mask cube and labelled as true positives. Real sources in the output catalogues of the three methods were differentiated from false positives by checking for the presence of an optical counterpart in the Panoramic Survey Telescope and Rapid Response System (Pan-STARRS) survey (\citealt{chambers2019panstarrs1}; see Sect. \ref{sec:real_sources}). Real sources were labelled as such and sources in the output catalogue not labelled as mock galaxies and that were without an optical counterpart were labelled as false positives. The masks of the identified real sources were created by taking the union of all the masks from the catalogues of the three source-finding methods. The masks for the real sources were then added to the ground truth and used to train V-Net's network, now with both real sources and mock galaxies in the training pointing. The newly trained network was then run on the test pointing and evaluated in comparison to SoFiA and MTO, using the ground truth containing both real sources and mock galaxies.

In an attempt to improve the purity of the source catalogues produced by all three methods, a post-processing step was implemented to remove as many false positive detections as possible. The rejection of false detections could have been based directly on the astrophysical plausibility of the attributes, such as the size of the galaxy. However, for large volumes like the Apertif survey cubes analysed here, many false positive detections have attributes very close to the attributes of sources. Therefore, a shallow machine learning approach was used instead, post-processing the source-finding catalogues with a random forest classifier \citep{598994}, as it is one of the best performing algorithms for classification \citep{RF_best}.

All three source-finding methods were run on both Apertif pointings. The three resulting catalogues of detected sources from the Apertif training pointing were combined and used as a single catalogue to train the random forest classifier, taken from scikit-learn \citep{scikit-learn}. This single catalogue was used to create a model that was generalisable to all three different source-finding methods. Due to the over-representation of false positive sources compared to true positive detections, the training catalogue had the potential to be imbalanced. In order to prevent this from creating a bias in the random forest classifier, the weights associated with each class, false positive detections and true positive sources, were automatically adjusted according to the frequency of that class by a factor of
\begin{align}
    \frac{N_{\text{detections}} }{N_{\text{classes}}N_{i}},
\end{align}

\noindent
where $N_{\text{detections}}$ is the number of detections in the catalogue, $N_{\text{classes}}$ is the number of classes, in this case being two, and $N_i$ is the number of detections in class $i$. The Gini index \citep{giniindex} was chosen to measure the quality of the splits in the decision trees of the random forest because it yielded the best improvement in the purity of the catalogues.

The volume in voxels, the total and peak fluxes in Jy$\,$km$\,$s$^{-1}$, the velocity width in km$\,$s$^{-1}$, the spatial dimensions in kpc, the \hi mass, and the spatial elongation of each detection in the Apertif training pointing catalogue was then used to train the random forest classification model. Once trained, the classification model was run on the catalogues of the Apertif test pointing, and the detections classified by the random forest classifier as false positive detections were removed from the catalogue to investigate if this improved the purity.

\section{Results}
\label{chap:results}

This section compares and evaluates the merits of all three source-finding methods, SoFiA, MTO, and the V-Net network, using their relative compute times and results.

\subsection{Compute times}
\label{sec:compute_times}

\begin{figure}
    \centering
    \includegraphics[width=\hsize]{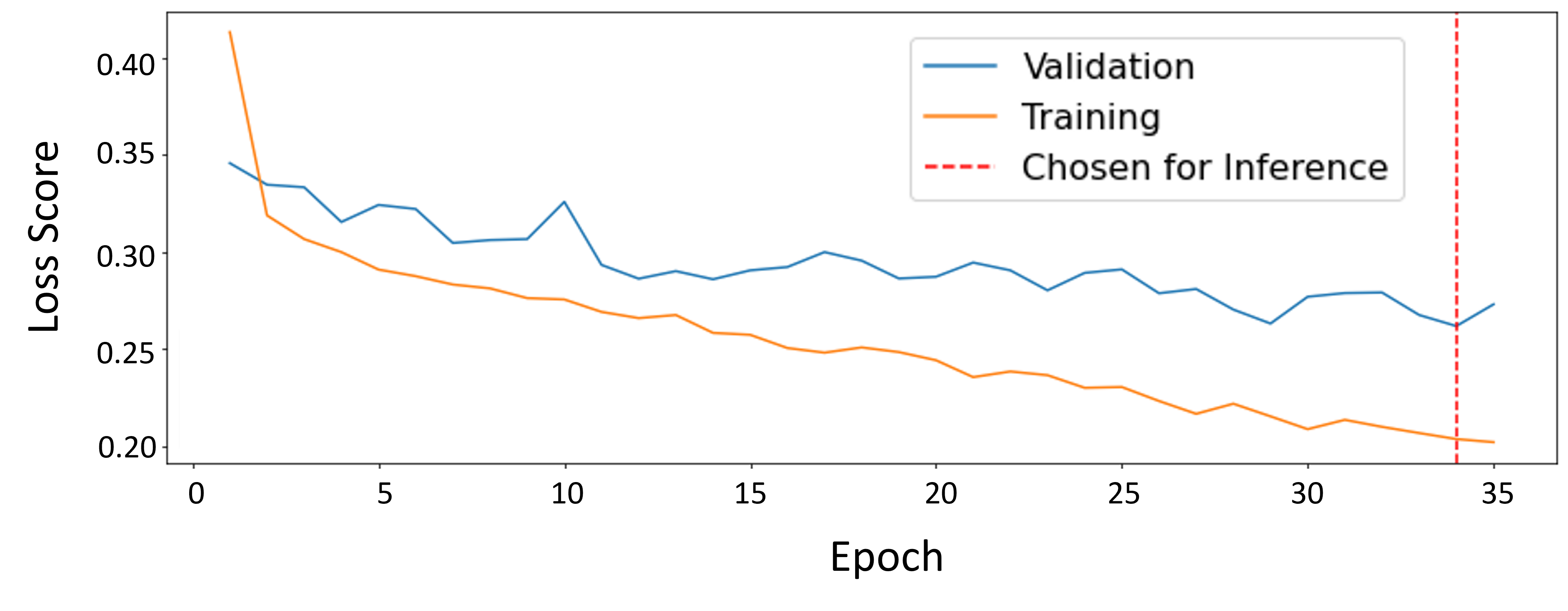}
    \caption{Training and validation loss scores of the V-Net network per epoch during the training process. The blue line shows the validation loss score and the orange line the training loss score, with the dashed red line indicating the point where the model was taken for inference.}
    \label{fig:exp1_training_loss}
\end{figure}

Before the V-Net network could be used as a source finder, the network was first trained using the masks of the mock galaxies and identified real sources as the ground truth. The learning progress of the network is illustrated in Fig. \ref{fig:exp1_training_loss}, showing both the training and validation loss scores as a function of the compute epoch on the Peregrine HPC cluster, as mentioned in Sect. \ref{chap:experiment}. About five epochs of the learning process could be completed per day. The trained network was taken from the 34$^{\mathrm{th}}$ epoch, to be used as a source-finding method. The trained network was taken from the 34$^{\mathrm{th}}$ epoch, to be used as a source-finding method. This epoch achieved the lowest combination of the training and validation loss score, obtained within the time frame of a week. It should be noted, however, that the time required to train the V-Net network is specific to the Peregrine cluster used, which was the most capable machine at our disposal.

After the training of the V-Net network, it was applied to the Apertif test pointing on the same machine as the MTO and SoFiA software. The V-Net network was found to be the slowest, taking just over twice as long as MTO to complete and more than 12 times longer than SoFiA. Again, it should be noted that these relative compute times depend on the parameter settings of the source-finding algorithms as listed in the Appendix. For instance, reducing or increasing the number of smoothing kernels for SoFiA will significantly affect the time it takes SoFiA to complete the task. The number and sizes of the smoothing kernels used by SoFiA, however, were deemed a good match to typical sizes of the sources in the Apertif cubes.

These relative compute times to completion were somewhat expected. While SoFiA could load an entire Apertif cube into memory, MTO had to work on sub-cubes sequentially, stitching together the masks of the detections from each sub-cube. The V-Net network had to work sequentially on even smaller sub-cubes than MTO due to the design of its architecture. It is important to recall that this evaluation of relative compute times only considers the process of source-finding for all three approaches. It does not include the time that was required to train the V-Net network.

\begin{table}
    \centering
    \caption{Elapsed real time in seconds for each of the three source-finding methods. \label{tab:perf}}
    \begin{tabular}{lc} \\ \hline 
    Method & Elapsed Time (minutes) \\ \hline \hline 
    SoFiA & 4.63 \\
    MTO & 26.97 \\
    V-Net  & 56.99 \\ \hline
\multicolumn{2}{l}{\footnotesize Note: The time is measured for a single spectral window}\\
\multicolumn{2}{l}{\footnotesize of size 1.0432 Giga voxels or $652\times1000\times1600$ voxels.}
    \end{tabular}
\end{table}

\subsection{Identifying real sources}
\label{sec:real_sources}

During the evaluation of these methods, the location of the brightest voxel (see Sect. \ref{sec:eval}) for each of the detected sources was used to cross-reference it with the ground truth masks of the mock galaxies. This allowed us to mark the detected sources as true or false positive detections accordingly. As previously mentioned, the collection of ground truth masks based on the inserted mock galaxies is incomplete, as it does not include the real sources that also exist within the cubes. Therefore, the false positives returned by each method were inspected by eye, by overlaying their moment-0 \hi emission maps as contours on optical images from Pan-STARRS \citep{chambers2019panstarrs1}. It should be noted that the 2D moment-0 maps were created by integrating the \hi emission over all the channels within the 3D masks generated by each source-finding method. It is important to emphasise here that each of the three methods may generate a different mask, and therefore a different moment-0 map, for the same detected source.

To reduce the effort of visually inspecting the false positive detections, they were first filtered according to their spatial and spectral extent to ensure only astrophysically plausible \hi sources were cross-referenced with the Pan-STARRS images. Only sources with spatial sizes smaller than 0.3$\,$Mpc, and with spectral widths between 7$\,$km$\,$s$^{-1}$ and 750$\,$km$\,$s$^{-1}$ were considered for visual inspection. Only detected sources that appeared to have an optical counterpart or to be realistic \hi emission sources were subsequently labelled as real sources, taking the union of the detected masks from each source-finding method as the ground truth mask for these real sources.

\begin{figure}
    \centering
    \includegraphics[width=\hsize]{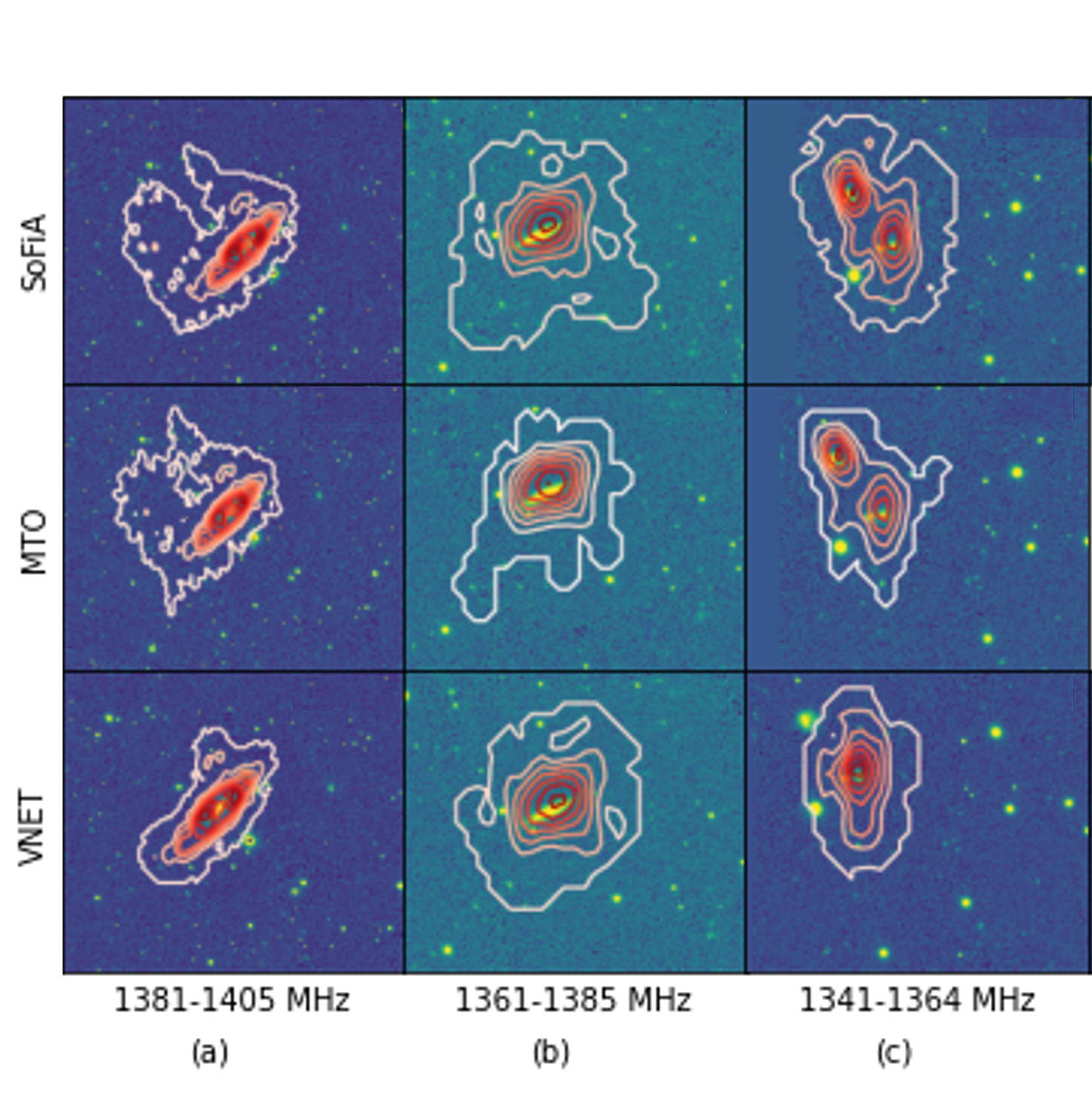}
    \caption{Examples of masked \hi moment-0 maps of detections of real sources, overlaid on their optical counterparts for each source-finding method. From left to right we show an example from different spectral windows of the Apertif test pointing with, from top to bottom, the mask from SoFiA, MTO, and the V-Net network. All the examples were matched across the methods by their brightest voxel.}
    \label{fig:real_examples}
\end{figure}

Figure \ref{fig:real_examples} presents examples of contoured \hi moment-0 maps overlaid on optical images of the same area of the sky for three real sources from three different spectral windows. In these examples, the real sources appear to show some asymmetry in their moment-0 maps as generated by SoFiA and MTO. For instance, considering the top and middle rows of Fig. \ref{fig:real_examples}, the moment-0 maps of source (a) appear to show signs of the \hi gas being stripped from the galaxy by ram pressure stripping, while source (c) appears to consist of two galaxies that are possibly interacting. However, these asymmetries are not recovered in the moment-0 maps generated by the V-Net network, as illustrated by the images in the bottom row of Fig. \ref{fig:real_examples}. This demonstrates that the V-Net network struggles to properly mask intrinsically asymmetric sources, which is expected as the network was primarily trained on symmetric mock galaxies and only the few identified real sources.

\begin{figure*}[ht]
    \centering
    \includegraphics[width=\hsize]{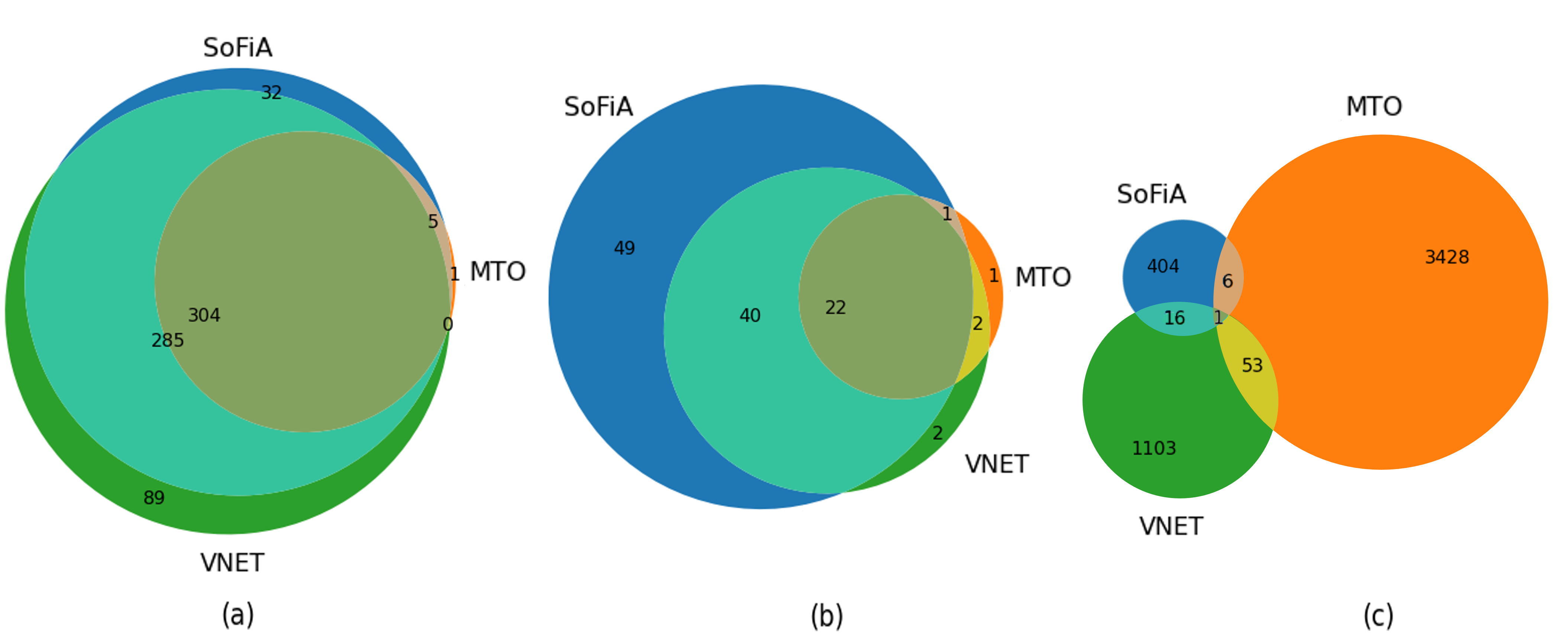}
    \caption{Venn diagrams of the total number of each kind of detection for all six spectral windows of the Apertif test pointing by each source-finding method prior to applying the random forest classifier. (a) Number of detected mock galaxies. (b) Number of detected real sources. (c) Number of detected false positives. SoFiA is represented by the blue circle, MTO is represented by the orange circle, and the V-Net network is represented by the green circle.}
    \label{fig:venn_exp1_test1}
\end{figure*}

The number of mock galaxies, real sources and false positives as detected by each method can be found in Table \ref{tab:detections}. We note that V-Net performs significantly better at locating mock galaxies in the training pointing than in the test pointing. This is expected, since the network was trained with the training pointing and was not exposed to the test pointing until it was tested. In order to visually summarise these results, Fig. \ref{fig:venn_exp1_test1} shows Venn \citep{venn} diagrams with the total number of detections from all six spectral windows of the Apertif test pointing.

\begin{table*}[ht]
    \centering
    \caption{Number of mock galaxies, real sources, and false positives detected by each source-finding method prior to applying the random forest classifier.}
    \label{tab:detections}
\begin{tabular}{ccccc|ccc|ccc}
\hline 
& Frequency &  & Mock & & & Real & &&  False &\\
& Range (MHz) &  & Galaxies & & & Sources & &&  Positives &\\ \hline \hline 
& & SoFiA & MTO & V-Net & SoFiA & MTO & V-Net & SoFiA & MTO & V-Net \\
Training  & 1280-1304 & 0 (0\%) & 2 (1\%) & 40 (13\%)  &  0 & 1 &  1 & 33 & 2069 & 44  \\
Pointing & 1300-1324 &  15 (5\%) & 1 (0\%) & 93 (31\%)  & 10 & 0 &  11 & 74 & 0 &  34  \\
& 1320-1344 &  54 (18\%) & 6 (2\%) & 145 (48\%)  &  3 & 0 & 2 & 52 & 1 &  35   \\
& 1341-1364 &  85 (28\%) & 20 (7\%) & 199 (66\%)   & 15 & 3 &  14 & 32 & 4 &  25 \\
& 1361-1385 & 175 (58\%) & 66 (22\%) & 258 (86\%)  & 10 & 3 & 8 & 156 & 10 &  13  \\
& 1381-1405 & 268 (89\%) & 190 (63\%) & 292 (97\%) & 41 & 21 &  35 & 70 & 9 &  12  \\
& \textbf{Total} & \textbf{597} & \textbf{285} & \textbf{1027} &  \textbf{79} & \textbf{28} &  \textbf{71} & \textbf{417} & \textbf{2093} &  \textbf{163}\\
& && &&&&&& &  \\
Test & 1280-1304 & 1 (0\%) & 1 (0\%) & 7 (2\%) & 0 & 0 & 0 & 23 & 3094 & 1030 \\
Pointing & 1300-1324 & 23 (8\%) & 0 (0\%) & 34 (11\%) & 23 & 1 &  11 & 184 & 13 &  20  \\
& 1320-1344 & 61 (20\%) & 5 (2\%) & 55 (18\%) &  1 & 0 & 0 & 65 & 12 &  21 \\
& 1341-1364 & 108 (36\%) & 26 (9\%) & 111 (37\%) &  40 & 4 &  20 & 79 & 65 &  40 \\
& 1361-1385 & 154 (51\%) & 79 (26\%) & 193 (64\%) & 24 &  7 &  17 &  13 & 64 &  33\\
& 1381-1405 & 279 (93\%) & 199 (66\%) & 278 (93\%) &  24 &  14 &  18 & 63 & 240 &  29\\
& \textbf{Total} & \textbf{626} & \textbf{310} & \textbf{678}&  \textbf{112} &  \textbf{26} &  \textbf{66} & \textbf{427} & \textbf{3488} &  \textbf{1173}\\  \hline
\multicolumn{11}{c}{\footnotesize Note: These detections are shown for each spectral window of the Apertif training and test pointings. For the mock}\\
\multicolumn{11}{c}{\footnotesize galaxies, the percentage of the total number of mock galaxies in the spectral window, 300, is shown in brackets.}
\end{tabular}
\end{table*}

Considering the detections from the test pointing, SoFiA appears to be the most successful at finding the real sources, with 49 real sources exclusively detected by SoFiA (Fig. \ref{fig:venn_exp1_test1} a). Comparing the physical properties of these 49 real sources to those of the real sources detected by all three source finders, it appears that SoFiA is better able to find sources that are spatially unresolved. On the contrary, the real sources detected by all three source-finding methods remain within a more conservative range for their spatial and spectral sizes. The V-Net network succeeds best at finding the mock galaxies, with 89 exclusively picked up by the V-Net network (Fig. \ref{fig:venn_exp1_test1} b).

MTO picks up the most false positives, with as many as 3428 false positives that neither SoFiA nor V-Net picked up. However, Table \ref{tab:detections} demonstrates that 90\% of these false positive detections are taken from the most distant spectral window. Similarly, around 88\% of the false positives detected by the V-Net network are located in this spectral window. This suggests that SoFiA's local noise scaling provides it with a significant advantage over the other source-finding methods in data cubes with higher RMS noise (see Table \ref{tab:apertif_properties}). The false positives located in the other five spectral windows show no clear difference in their properties when comparing the detections of SoFiA and V-Net, but MTO appears to be more vulnerable to artefacts with spectral widths below 25 km$\,$s$^{-1}$.

It is relevant to note here that the different source-finding methods respond differently to deviations from uniform noise. For example, due to radio frequency interference (RFI) that was not fully flagged, due to an imperfect subtraction of astrophysical radio continuum sources or due to calibration errors. For instance, SoFiA is sensitive to very local enhancements of the noise that could not be adequately normalised, as may happen at locations of poorly subtracted continuum sources or near the edges of the spectral windows. MTO, on the other hand, tends to pick up more extended artefacts and sharp-edged geometric artefacts related to calibration errors. V-Net appears to favour artefacts similar to those picked up by MTO, but with more elongation, such as ripples due to RFI residuals. Each method, therefore, has its own shortcomings, which is expected due to the differences in their approaches.

\subsection{Evaluating the source finders}
\label{sec:eval}

In order to compare and evaluate each of the source-finding methods in each experiment, the resulting 3D masks, as produced by each source-finding method, were mapped one-to-one to the ground truth masks for each mock galaxy using the brightest voxel of each source. If a mask from a detection contained the brightest voxel of more than one source within the ground truth masks, the source with the voxel of the highest flux was chosen. Two metrics were used to measure how well the detections of a source finder masked the sources. The under-merging metric, UM, measures the extent to which a masking method breaks up sources that should be contained within single masks. The over-masking metric, OM, measures the extent to which a masking method masks more than just the source. Both metrics are defined by \citet{Haigh_2021}, who adapted the work of \citet{levine1981experimental}, and are extended to three dimensions as follows:

\begin{align}
    \mathrm{UM} &= \sum^N_{k=1}\frac{(V_k-(T_k\cap R_k))(T_k\cap R_k)}{V_k}, \\
    \mathrm{OM} &= \sum^N_{k=1}\frac{(v_k-(T_k\cap R_k))(T_k\cap R_k)}{v_k},
\end{align}

\noindent
where $N$ is the number of one-to-one mapped detections to a ground truth mask. $R_k$ is the ground truth mask, with volume $V_k$, and $T_k$ is the mask of the detected source, with volume $v_k$. $T_k\cap R_k$ is, therefore, the intersection of these two masks. The combination of the two metrics results in the volume score, which indicates the quality of the mask of a detected source. The volume score is defined as follows \citep{Haigh_2021}:

\begin{align}
    \mathrm{V}_{\mathrm{score}} &= 1-\sqrt{\mathrm{OM}^2+\mathrm{UM}^2}.
\end{align}

\noindent
A final combined score was used and the mean of this value for each of the six spectral windows was calculated to rank and compare all three source-finding methods. This score takes the purity and completeness of the source finders into account by using the $F_{\mathrm{score}}$ from Eq. \ref{eq:fscore}, as well as the merging scores as follows:

\begin{align}
    \mathrm{C}_{\mathrm{score}} &= \sqrt[3]{(1-\mathrm{OM}) (1-\mathrm{UM}) F_{\mathrm{score}}} .
\end{align}

\begin{figure*}[ht]
    \centering
    \includegraphics[width=\hsize]{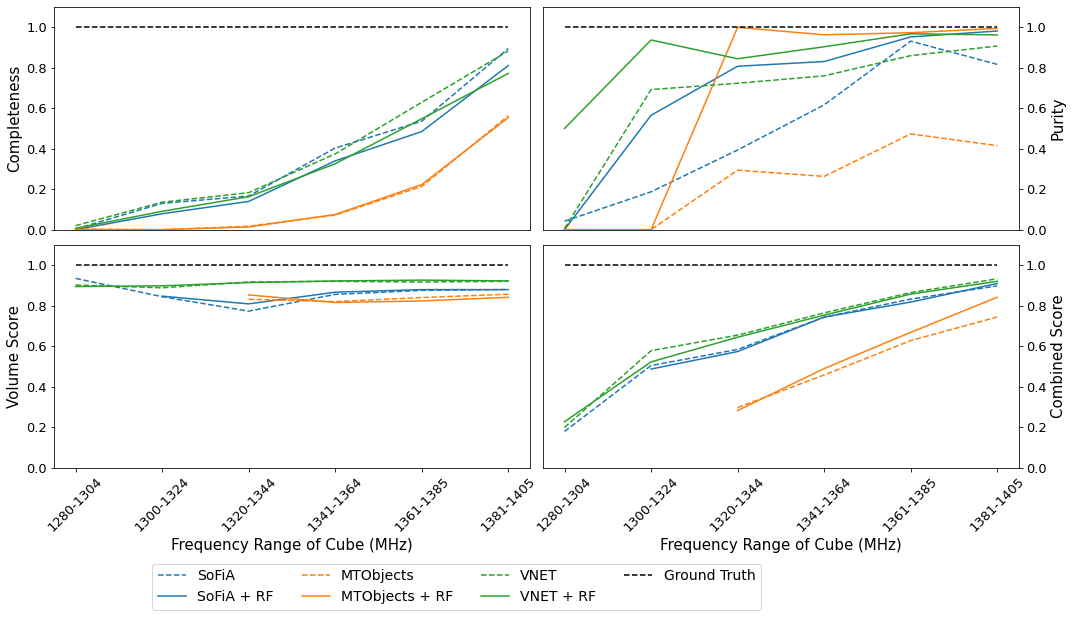}
    \caption{Various evaluation metrics of the three source finders evaluated with both the mock galaxies and the real sources as a function of the frequency range corresponding to each spectral window. For each plot, the results of SoFiA are shown in blue, the results of MTO in orange, and the results of the V-Net network in green, and the ground truth is shown as a dashed black line. The panel at the top left shows the completeness and the top right shows the purity. The mask quality is shown with the volume score in the bottom-left plot, and the combination of all merging scores, purity, and completeness is shown with the combined score in the bottom-right plot. For all the plots the x-axis shows the frequency range associated with each of the spectral windows of the Apertif test pointing.}
    \label{fig:completeness_test_exp1}
\end{figure*}

\noindent
These evaluation metrics, for each source finder at each frequency range, can be found in Fig. \ref{fig:completeness_test_exp1}, represented by the dashed lines. From these plots, it is clear that all the methods improve both their completeness and purity with frequency, as the spectral windows are closer to the observer and therefore contain seemingly brighter and larger sources. In addition, all the methods have very high volume scores above 0.8, demonstrating the high mask quality of them all.

When comparing the methods, the V-Net network appears to be the best performing method, with the most consistent purity across the different spectral windows. MTO, on the other hand, while a close competitor with the other methods in its mask quality, indicated by the volume score, seems to struggle most with its completeness and purity. This is most likely due to the way in which the noise was modelled, allowing MTO to identify the noise as sources. The addition of the random forest classifier, to improve the purity of all source-finding methods, is indicated in Fig. \ref{fig:completeness_test_exp1} by the solid lines. While this largely improves the purity, seen in the upper right plot, this comes at the expense of the completeness for all methods but MTO, seen in the upper-left plot.

Despite having a built-in reliability filter, the results indicate that the random forest classifier also succeeds in improving the purity of SoFiA's catalogue. Comparing the properties of the false positives detected by SoFiA but rejected by the random forest classifier to those that are selected by the random forest classifier, the largest difference is the distribution of the mask volumes. The number of false positive detections rejected by the random forest suddenly increases for a mask volume below 40 voxels. Since the reliability measurements of SoFiA are calculated using only the values of the voxels within the mask, these results suggest the reliability algorithm could better filter unreliable sources by using the volume and shapes of the masks as additional parameters.

\subsection{Comparing the physical properties}

The physical properties of the masks created by each source-finding method were calculated to compare and search for any significant differences between the methods. The spatial sizes of the masks, $\Delta x$ and $\Delta y$, as well as the velocity width, $\Delta v$, were calculated following \citet{1705042146:online}. In addition to the physical size of the object, the spatial elongation of each mask was calculated following the description in Sect. \ref{sec:MTO}. The peak and total fluxes, and the \hi mass were calculated following \citet{1705042146:online}.

\begin{figure}
    \centering
    \includegraphics[width=\hsize]{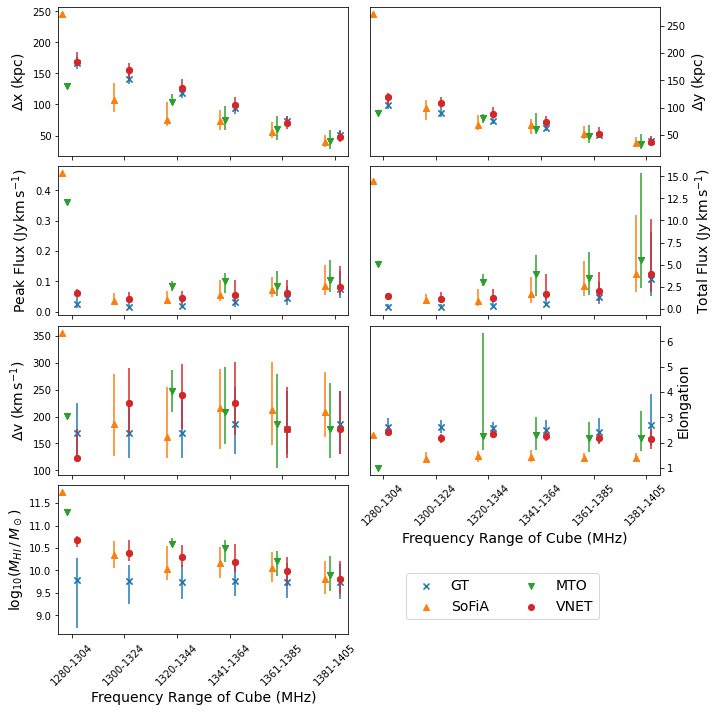}
    \caption{25th, 50th, and 75th percentiles of the physical properties of the masked mock galaxies detected by each method per spectral window. For all the plots the x-axis shows the frequency range associated with each spectral window, and the markers indicate the median or 50th percentile. The orange triangles show the detections of SoFiA, the green triangles show the detections of MTO, the red circles show the detections of the V-Net network, and the blue crosses show the ground truth masks. The lines below and above each marker indicate the 25th and 75th percentiles, respectively. The top panel shows the two spatial dimensions, $\Delta$x and $\Delta$y, in kpc, the second panel shows the peak flux and the total flux, the third panel shows the velocity width, $\Delta$v, in km/s and the spatial elongation, and the last panel shows the log of the \hi mass.}
    \label{fig:exp1_mask_properties}
\end{figure}

The 25th, 50th, and 75th percentiles of the physical properties of the masked mock galaxies were investigated in order to further compare the quality of the masks. In Fig. \ref{fig:exp1_mask_properties} the spatial dimensions and the total flux of the masked mock galaxies detected by all the methods appear to be almost in agreement, only slightly veering away from each other and the ground truth at the lower frequencies. While SoFiA appears to be the least accurate at masking the spatial sizes and elongation of the mock galaxies, the V-Net network is the most accurate. However, this is only true for the mock galaxies, on which the V-Net network was trained since Fig. \ref{fig:real_examples} shows that it is less accurate at masking less regular real sources.
\section{Discussion}
\label{chap:discussion}
In the search for the optimal pipeline to locate and mask \hi sources in a 3D \hi emission cube, three source-finding methods were compared, namely SoFiA, MTO, and a 3D CNN using the V-Net architecture. In an attempt to improve the purity of these methods, a random forest classifier was then used to post-process the resulting catalogues.

\subsection{Which is the optimal source-finding pipeline?}

For a final comparison of all the source-finding pipelines proposed, the mean combined score across the spectral windows was calculated for each. The resulting scores can be found in Table \ref{tab:final_scores}. Evaluating the three source-finding methods on their own, the V-Net network is found to rank best overall, shown by the `Total' column. However, the overall score does not account for the fact that there are far fewer real sources than mock galaxies in the data set and so these detections were then evaluated separately. SoFiA is found to rank the best regarding real source detections, while the V-Net network ranks the best at detecting mock galaxies. To account for this imbalance, the mean of the `Real' and `Mock' scores was calculated, resulting in SoFiA as the optimal source finder.
\begin{table}
    \centering
    \caption{Mean combined scores of all source-finding pipelines. \label{tab:final_scores}}
\begin{tabular}{lcccc}
\hline
& Real & Mock & Total & Mean \\ \hline \hline
SoFiA $+$ random forest & \textbf{0.689} & 0.684 & 0.730 & \textbf{0.686} \\
SoFiA & 0.651 & 0.703 & 0.747 & 0.677 \\
V-Net $+$ random forest & 0.472 & \textbf{0.763} & \textbf{0.773} & 0.610 \\
V-Net & 0.449 & 0.752 & 0.764 & 0.600 \\
MTO $+$ random forest & 0.317 & 0.553 & 0.563 & 0.435 \\
MTO & 0.201 & 0.421 & 0.420 & 0.311 \\ \hline
\multicolumn{5}{l}{\footnotesize Note: The `Total' score represents the pipeline's ability to find all} \\
\multicolumn{5}{l}{\footnotesize the real sources and mock galaxies together, while the `Mean'} \\
\multicolumn{5}{l}{\footnotesize score represents the mean of the `Mock' and `Real' scores, which} \\
\multicolumn{5}{l}{\footnotesize evaluate the detections of mock galaxies and real sources separately.} \\
\multicolumn{5}{l}{\footnotesize The pipelines with $+$ random forest indicate that a random forest} \\
\multicolumn{5}{l}{\footnotesize classifier was used to post-process the results.}
\end{tabular}
\end{table}

Since SoFiA is the most widely used and optimised source-finding software for \hi emission data, it makes sense that of all the source-finding methods it is most suited to the data set used. This is perhaps due to its ability to perform local noise scaling, as well as its reliability filter, which reduces the false detection of noise and artefacts. In addition, the struggle of the V-Net network to mask real sources is likely due to the fact that it was trained with majority symmetric mock galaxies and could be overfitting to them \citep{DBLP}.

Of all the methods, SoFiA also takes the least amount of compute time, while the V-Net network was found to be the slowest (see Table \ref{tab:perf}). It is also important to point out the large number of computational resources used and the time taken for the training of the V-Net network. Therefore, it is suggested that in future work the sliding window is applied in parallel or the architecture is modified to take larger inputs since both would reduce the time spent on the training and source-finding.

However, investigating the physical properties of the masked mock galaxies detected, the V-Net network is found to yield the closest values to the ground truth. Outperforming the other source-finding methods at masking the mock galaxies suggests the potential for the V-Net network to outperform SoFiA overall, with an even better mask quality, should it be trained on more real sources. Alternatively, to remove the difficulty of not having enough labelled real sources, there is potential to investigate an unsupervised deep learning solution. It is also important to note the advantage of the V-Net network is that it is less sensitive to input parameters than the other source-finding methods.
 
Although MTO struggles with a low purity on its own, this was expected since \citet{Arnoldus2015} used the reliability filter from SoFiA to tackle this and remove artefacts and noise detections. This low purity is most likely due to the way in which the background is estimated, which is not suitable for radio data with a mean near zero. This noise modelling could be the cause of the struggle of MTO to differentiate sources from the noise, which is suggested by the fact that 90\% of the false positive detections are located in the spectral window with the highest RMS noise. While the completeness of MTO was not necessarily expected to overtake SoFiA, it is not as competitive in this regard as found in \citet{Arnoldus2015}, reaching up to 30\% difference at some spectral windows when compared to SoFiA. While this could be simply due to the different sets of data used, it could also be caused by the lack of spectral smoothing, since only the spatial dimensions were smoothed anisotropically during the pre-processing.

As \citet{Arnoldus2015} suggests, this could perhaps be improved with some experimentation with spectral smoothing, including the use of isotropic smoothing. Since the mask quality of MTO is very competitive with the other methods, improving the purity and completeness could also potentially enable it to overtake as the optimal source-finding method.

\subsection{Can the addition of a classical machine learning classifier improve the results of the source finders?}
\label{random forestclassifier}

The addition of a random forest classifier as a post-processor of the source-finding catalogues is found to improve the purity of all the source-finding methods. Besides MTO, it also causes a slight drop in completeness. This results in the overall improvement, measured by the combined score, for all the source finders. This improves the purity of MTO further than the reliability filter used in \citet{Arnoldus2015}, such that it outperforms the other source-finding methods in terms of purity, with little to no effect on the completeness. This suggests that implementing the random forest classifier within the max-tree could make the purity of MTO competitive on its own.

The top features ranked according to the random forest classifier are found to be the velocity width and the \hi mass, demonstrating the importance of the physical size and flux of the detections for their reality. This could help inform the filtering process in rule-based methods like MTO. In addition, the purity of the SoFiA catalogue is also improved with the addition of the random forest classifier. This suggests that taking the shape and volume of detected masks into account could improve the reliability algorithm of SoFiA. However, it would also be interesting in future work to compare the results of SoFiA with and without the reliability filter to better understand how the random forest classifier improves the purity of the catalogue.

\subsection{What are the limitations of these results?}

The locating of real sources in the \hi emission cubes enabled a separate evaluation of the methods with both real sources and mock galaxies. However, manually finding the real sources during the cross-matching process was very time-consuming and still resulted in an over-representation of the mock galaxies relative to the real sources. This indicates that this approach to labelling the ground truth was not the most effective. The over-representation of mock galaxies not only biased the training of the V-Net network but also could have biased the evaluation of all the pipelines for more symmetric, bright sources.

An additional limitation of the data used was the partial labelling of the data set since there is no guarantee that all the real sources existing prior to the mock galaxy insertion were labelled. The limitations of manually optimising the parameters for SoFiA and MTO must also be considered. Not only is this limited as a subjective approach, but identifying false positives can be a prohibitive process. We, therefore, suggest that further works make use of automatic optimisation algorithms when comparing different source-finding methods.
\section{Conclusion}
\label{chap:conclusion}
Due to improvements in technology, the amount of data coming from astronomical surveys continues to increase, and therefore so does the need for fast and accurate techniques for detecting and characterising sources. The challenge lies in the lack of clarity in the boundaries of sources, with many having intensities very close to the noise. This project, therefore, aimed to find the best pipeline for finding and masking the most sources with the best mask quality and the fewest artefacts in 3D neutral hydrogen cubes. This was done by testing two existing statistical methods, SoFiA and MTObjects, and a well-known medical imaging 3D CNN architecture called V-Net on \hi emission data taken from the Apertif medium-deep survey with inserted mock galaxies.

The highest performing source finder of the methods tested is SoFiA. This is due to this pipeline's ability to best find and mask real sources, as well as the small amount of time and computational resources needed relative to the other methods. However, the V-Net network ranks a close second, thanks to its ability to detect mock galaxies better than real sources. The addition of a random forest classifier is found to improve the purity of all the source-finding methods. SoFiA could therefore benefit from adding a built-in random forest classifier to improve the ease of use of this post-processing step.

As the everyday applications of computer vision continue to increase, so does the amount of astronomical data produced by ever-improving telescopes. With this progress grows the need for astronomers to keep up to date with the advances in computer vision. In this paper, a computer vision solution was applied to an astronomical task and found to be competitive with traditional software. In addition, with the implementation of suggested improvements, it is likely to even overtake the traditional astronomy methods. This demonstrates that the interchangeability of solutions between the two fields will become fundamental to the growth of data analysis in future astronomy.

\begin{acknowledgements}
      We thank the Center for Information Technology of the University of Groningen for their support and for providing access to the Peregrine high-performance computing cluster. This work was funded by the Nederlandse Onderzoekschool Voor Astronomie (NOVA).
\end{acknowledgements}
\bibliographystyle{agsm}
\bibliography{references}

\newpage

\begin{appendix}

\section{Parameters of the source-finding methods}
\begin{table}[H]
    \centering
    \caption{Chosen parameters for SoFiA when applied to the nearest spectral window. \label{tab:sofia_parameters}}
\begin{tabular}{cc}
\hline 
Parameter Name & Set Value \\ \hline \hline 
scaleNoise.windowXY & 75 \\
scaleNoise.windowZ & 51 \\
threshold.fluxRange & full \\
scfind.enable & true \\
scfind.kernelsXY & 0, 3.6, 6.6 \\
scfind.kernelsZ & 0, 3, 5, 9, 15 \\
scfind.threshold & 3.5 \\
linker.enable & true \\
linker.maxSizeZ & 133 \\
linker.maxSizeXY & 0 \\
linker.minSizeZ & 3 \\
linker.minSizeXY & 2 \\
reliability.scaleKernel & 0.5 \\
reliability.threshold & 0.9 \\
reliability.minSNR & 3 \\ \hline
\multicolumn{2}{l}{\footnotesize Note: A full description of the SoFiA 2.3.1 control parameters}\\
\multicolumn{2}{l}{\footnotesize  and their defaults   \citep{sofia_2}, which were taken for the}\\
\multicolumn{2}{l}{\footnotesize unmentioned parameters in this table, can be found in the}\\
\multicolumn{2}{l}{\footnotesize \href{https://github.com/SoFiA-Admin/SoFiA-2/wiki/SoFiA-2-Control-Parameters}{SoFiA repository}.}
\end{tabular}
\end{table}

\begin{table}[H]
    \centering
    \caption[The chosen parameters for MTObjects]{Chosen parameters for MTObjects.}
\begin{tabular}{cc}
\hline 
Parameter Name & Set Value \\ \hline \hline 
Threads & 64 \\
Connectivity & 6 \\
MeanBackground & Dependent on window \\
StdDev & Dependent on window \\
MoveupFactor & 0.2 \\
BackgroundEstimation & True \\
ShowObjects & True \\
InverseGain & 0 \\
Lambda & 0 \\
BitsPerPixel & 32 \\
SignificanceLevel & 1$\times$10$^{-5}$ \\
ExportAttributes & False \\
StatisticalTest  & "FluxDensityRadio" \\
SmoothingDegree & 0 \\
MoveWrtIntensity & False \\
DetectNestedObjects & True \\
MinSpectralWidth & 3 \\
MinSpatialWidth & 2 \\
\hline
\multicolumn{2}{l}{\footnotesize Note: These parameters were used for the MTObjects}\\
\multicolumn{2}{l}{\footnotesize implemented by \citet{Arnoldus2015}. The mean (\textbf{MeanBackground})}\\
\multicolumn{2}{l}{\footnotesize  and standard deviation (\textbf{StdDev}) of the background were calculated}\\
\multicolumn{2}{l}{\footnotesize for each window of the sliding window. In addition, \textbf{InverseGain}}\\
\multicolumn{2}{l}{\footnotesize and \textbf{Lambda} are depreciated and therefore were not set above zero.}\\
\end{tabular}
\label{tab:mto_param}
\end{table}

\begin{table}[H]
    \centering
    \caption{Chosen parameters for V-Net.}
\begin{tabular}{cc}
\hline 
Parameter Name & Set Value \\ \hline \hline 
Optimizer & ADAM $^4$\\
Batch Size & 4 \\
Learning Rate & 1$\times$10$^{-3}$ \\
\hline
\multicolumn{2}{l}{\footnotesize Note: The V-Net model used was implemented by}\\
\multicolumn{2}{l}{\footnotesize \href{https://github.com/black0017/MedicalZooPytorch}{MedicalZooPytorch} \citep{adaloglou2019MRIsegmentation}.}\\
\multicolumn{2}{l}{\footnotesize $^4$ Adaptive moment estimation \citep{kingma2017adam}}
\end{tabular}
\label{tab:vnet_param}
\end{table}
\end{appendix}

\end{document}